# Net 582-Gb/s C-band and 4×526-Gb/s O-band IMDD Transmission Using Ultra-broadband InP-DHBT-based Electrical Mixer

Masanori Nakamura(1), Teruo Jyo(2), Munehiko Nagatani(1, 2), Hitoshi Wakita(2), Miwa Mutoh(2), Yuta Shiratori(2), Hiroki Taniguchi(1), Akira Masuda(1), Shuto Yamamoto(1), Fukutaro Hamaoka(1), Etsushi Yamazaki(1), Hiroyuki Takahashi(1, 2), Takayuki Kobayashi(1), and Yutaka Miyamoto(1)

(1) NTT Network Innovation Laboratories, NTT Corporation, 1-1 Hikari-no-oka, Yokosuka, Japan, msnr.nakamura@ntt.com
(2) NTT Device Technology Laboratories, NTT Corporation, 3-1 Morinosato Wakamiya, Atsugi, Japan

**Abstract** *We successfully transmitted a net 582-Gb/s probabilistically shaped PAM12 C-band signal over 11-km dispersion-shifted fibre and net 4×526-Gb/s uniform PAM8 O-band signals over 2-km four-core fibre using a single-carrier 216-GBd IMDD system based on a 150-GHz bandwidth InP-DHBT electrical mixer and a thin-film lithium-niobate modulator.* 

## Introduction

To cope with the rapid increase in network traffic, especially in data centre networks (DCNs), a high-speed intensity-modulated direct detection (IMDD) system is required as a cost-effective solution for short-reach applications used in next-generation Ethernet [1]. One way to reduce the cost per bit while accommodating the next generation Ethernet (e.g., 1.6 TbE and more) is to use a net bitrate over 400 Gb/s/lane with a high-speed signal beyond 200 GBd by reducing the number of lanes. Figure 1 shows recently reported high-speed IMDD experiments [2–11], where the highest net bitrate of 548.7 Gb/s/lane has been demonstrated [7] in the C-band. In the IMDD system, chromatic dispersion (CD) is the main limitation factor of transmission distance as the symbol rate increases. Extreme-performance high-speed IMDD transmission experiments for the C-band have been demonstrated mainly over several hundred meters of standard single-mode fibre (SSMF) [5], SSMF with dispersion compensating fibre [3, 4, 7], or dispersion shifted fibre (DSF) [8–10].

An SSMF has a zero-dispersion wavelength of around 1310 nm, corresponding to the O-band. Therefore, practical high-speed IMDD transmission systems over 2 km and beyond have mainly focused on O-band systems. Prior work has demonstrated 1.6 Tb/s (4×400 Gb/s/lane) transmission over 10-km installed four-core fibre using O-band 155-GBd pulse amplitude modulation 8-level (PAM8) signals [2]. A net 456 Gb/s O-band 190-GBd PAM8 signal transmission over 2-km SSMF was demonstrated using a thin-film lithium-niobate (TFLN) modulator [11]. However, at present, there have been no demonstrations of an O-band IMDD transmission with a net >500 Gb/s/lane using a symbol rate of >200 GBd.

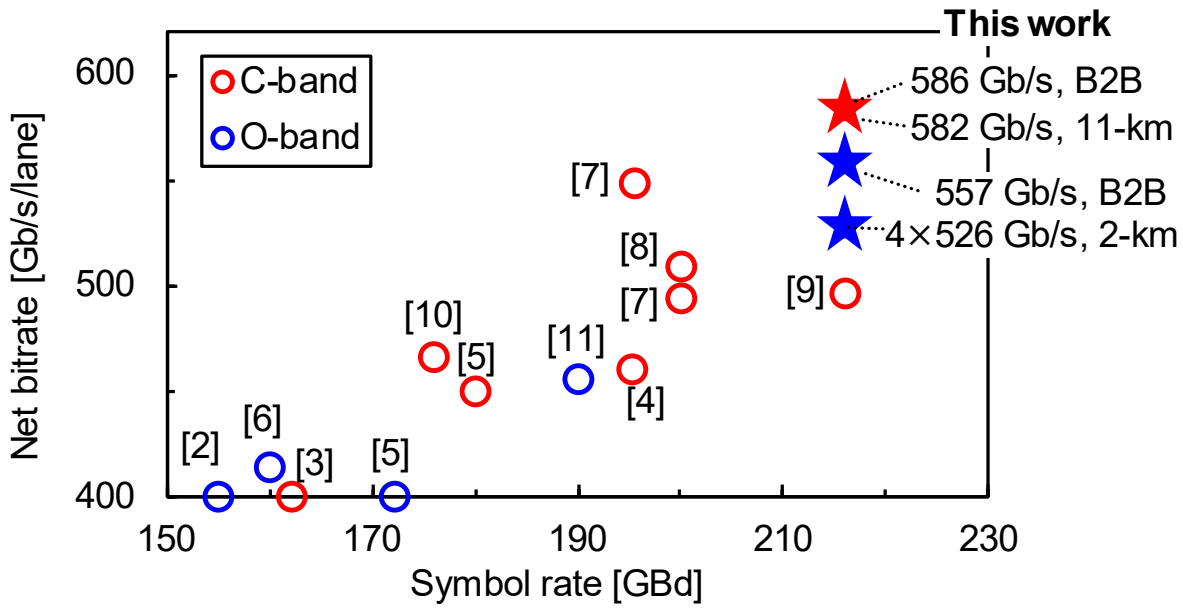

**Fig. 1:** High-speed IMDD experiments with a net bitrate per lane of ≥400 Gb/s and a symbol rate of ≥150 GBd.

In this paper, we demonstrate a record net bitrate of a net 582-Gb/s probabilistically shaped (PS) PAM12 signal over 11-km DSF in the C-band using a single-carrier 216-GBd IMDD transmission system based on a 150-GHz bandwidth (BW) indium phosphide double heterojunction bipolar transistor (InP-DHBT)-based electrical mixer and a thin-film lithium-niobate modulator to validate the high-speed signal generation >200 GBd and net >500 Gb/s signal transmission in the C-band. Then, using the broadband configuration, we are able to achieve a net 557 Gb/s O-band 216-GBd uniform PAM8 signal generation and detection at back-to-back configuration. Moreover, 2.1 Tb/s transmission with net 4×526-Gb/s uniform PAM8 signals over 2-km four-core fibre is demonstrated in the O-band. As can be seen in Fig. 1, we achieved the highest net bitrate per lane in both the C- and O-bands.

## Experimental setup

Figure 2(a) shows the experimental setup for a high-speed IMDD transmission over an 11-km DSF in the C-band and a 2-km four-core fibre in the O-band. The high-speed electrical signals were generated using our in-house active combiner [12] and a newly developed mixer, which have BWs of 150 GHz made possible by our InP-DHBT technology [13]. The mixer module shown in Fig. 2(b) has an intermediate frequency (IF) port with a 1-mm connector, a radio frequency (RF) port with a 0.8-mm connector, and local oscillator (LO) ports consisting of positive (p) and negative (n) ports with a 1-mm connector. As shown in Fig. 2(c), the mixer has a 150-GHz BW in the up-conversion gain characteristics at a 75-GHz LO configuration. The high-speed signal generation with digital pre-processing [14] was implemented for our configuration using the broadband mixer and the active combiner, as follows. A low-pass and high-pass filter (LPF and HPF) in a Tx-DSP divided the target wideband signal into lower- and upper-band signals. The upper-band signal was digitally down-converted with the LO frequency. These pre-processed signals were output from 256-GSa/s arbitrary waveform generator (AWG) with a BW of 80 GHz. The lower-band signal was input to the mixer and up-converted with the LOs. The AWG also generated clock signals for the LOs, where analogue bandpass filters (BPFs) were used to eliminate unwanted spurious tones. The analogue HPF cut the lower-side frequency image from the mixer and IF and LO leakages. Finally, the active combiner added the lower- and upper-band signals to reconstruct the desired wideband signal. In the C-band and O-band experiments, the analogue HPF cutoff frequency was 75 and 82 GHz, the clock frequency for LOs was set to 72 and 76 GHz, and the cutoff frequency of digital LPF and HPF was set to 76 and 82 GHz, respectively. A 130-GHz-BW amplifier [15] was added between the HFP and the active combiner in the O-band experiments to increase the symbol rate.

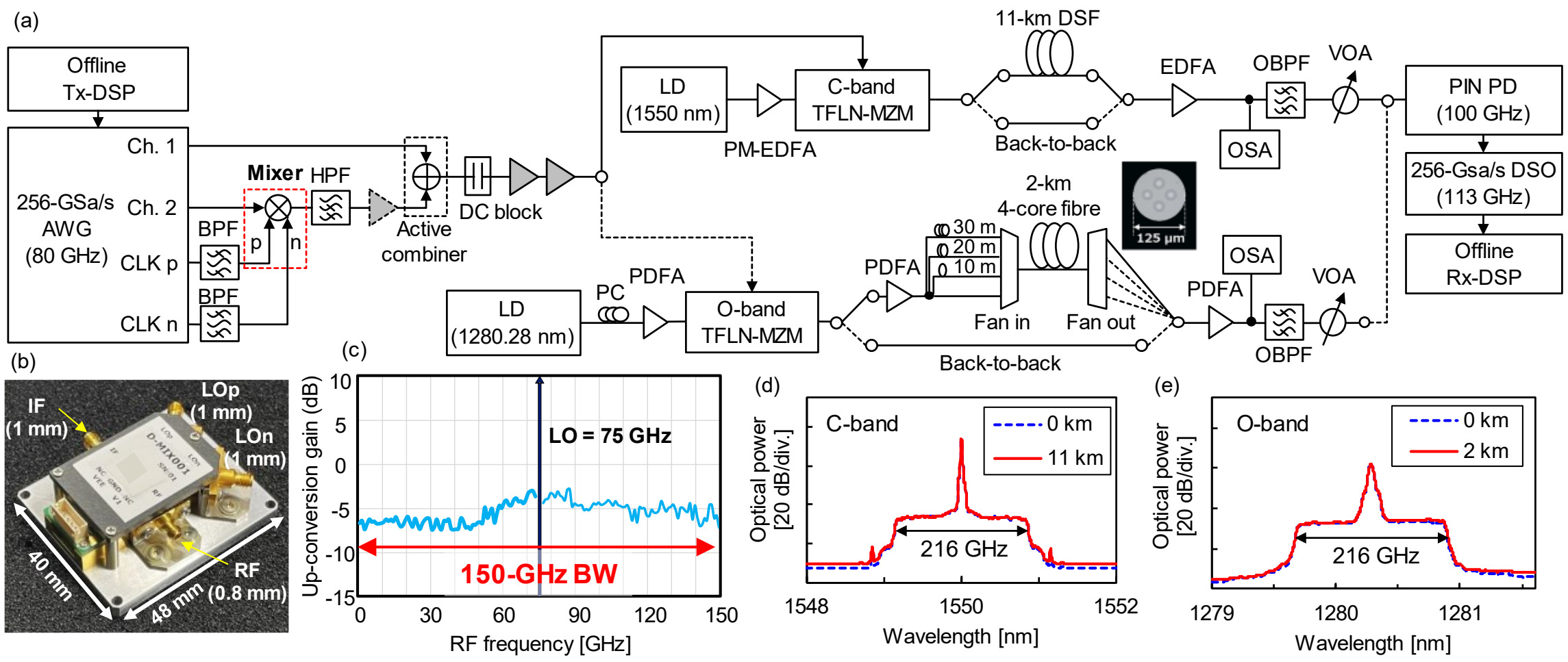

**Fig. 2:** (a) Experimental setup for high-speed IMDD transmission over 11-km DSF in C-band and 2-km four-core fibre in O-band. (b) 150-GHz-BW InP-DHBT mixer module. (c) Up-conversion gain characteristics of the mixer module. Optical spectra for 216-GBd signal before and after transmission in (d) C-band and (e) O-band.

In the Tx-DSP, PAM symbols with a sequence length of ~2.6×$10^5$ were generated from the Mersenne twister. To compensate for nonlinearities in the transmitter (Tx), we applied nonlinear digital pre-distortion based on a third-order Volterra filter with memories of 31 symbols [16]. The filter coefficients were learned using reference sequences generated with different seeds from the sequences for performance evaluation to avoid overfitting. The compensated symbol sequence was up-sampled using a root-raised cosine filter with a roll-off factor of 0.01. Then, the digital pre-processing described above was applied to the signal. The signals were input to the AWG after being digitally compensated for the frequency responses of the Tx devices.

The optical signal from the laser diodes (LDs) was amplified to an optical power of 20 dBm using a polarization-maintained (PM) erbium-doped fibre amplifier (EDFA) in the C-band and a praseodymium-doped fibre amplifier (PDFA) with a polarisation controller (PC) in the O-band. The optical signals were modulated using THLN Mach-Zehnder modulators (THLN-MZMs) with a BW over 110 GHz at 4.5 dB, which have a Vpi of 2.8 V and 2.5 V in the C- and O-bands, respectively. The wideband electrical signal was input to the MZM after being amplified by a 130-GHz-BW amplifier with a gain of ~7 dB [15] and a ~100-GHz-BW driver amplifier with a gain of ~16 dB.

The transmission lines in the C- and O-band experiments used an 11-km DSF and a 2-km uncoupled four-core fibre with a clad diameter of 125 µm, respectively. The zero-dispersion wavelength of each core was around 1280 nm. The input signals for the four-core fibre were de-correlated with delay lines. The transmitted signals were amplified, followed by optical BPF (OBPF) to cut amplified spontaneous emission noise. In the C-band experiments, the OBPF, consisting of a flexible grid wavelength selective switch, also compensated for the BW limitations of the PIN photodiode (PIN-PD) and residual CD. Figure 2(d) and (e) show the optical spectra for 216-GBd signals, measured by the optical spectrum analyser (OSA), in the back-to-back configuration and after transmission, before the OBPF. The received signal was detected by the 100-GHz-BW PIN-PD and digitalised by a 113-GHz-BW 256-GSa/s digital storage oscilloscope (DSO). The digital signal was equalised with a T/2-spaced feed-forward equaliser to recover the symbols. We calculated the log-likelihood ratio from the symbols to estimate the normalised generalised mutual information (NGMI) and the required code rate to achieve error-free decoding after soft-decision forward error correction (SD-FEC) in the same manner as [9]. A low-density parity check (LDPC) code defined by DVB-S2 [17] and puncturing-based rate adaptive coding were used to evaluate the code rate, assuming a 0.79% overhead outer hard decision FEC [18].

## Results and discussion

We first maximised the net bitrate in the C-band experiments. Figure 3 shows the achievable and net bitrate as a function of entropy of 216-GBd PS-PAM12 signals in the back-to-back configuration and after the 11-km DSF transmission. The PS-PAM12 signals, which followed a Maxwell-Boltzmann distribution, were generated by probabilistic amplitude shaping with a constant composition distribution matcher [19]. The achievable bitrate and net bitrate were respectively calculated from the NGMI and the required code rate as $C = \{H - (1 - R) \times 4\} \times 0.216$, where $C$ is the achievable bitrate or net bitrate for the PS-PAM12, and $R$ is the NGMI (for achievable bitrate) or the required code rate (for net bitrate). The highest net bitrate was 586.8 Gb/s at the back-to-back configuration and 582.6 Gb/s after 11-km transmission. The inset of Fig. 3(b) shows the received-symbol amplitude histogram with the highest net bitrate configuration of the PS-PAM12 signal after 11-km transmission.

Next, we maximised the net bitrate of uniform PAM8 signals in the O-band experiments using a simple configuration without an additional distribution matcher block by varying the symbol rate from 208 to 228 GBd in the back-to-back configuration. Figure 4(a) shows the NGMI, code rate, and bit error rate (BER) for each symbol rate. The achievable and net bitrate of the uniform PAM8 signal, as shown in Fig. 4(b), were calculated as $C = 3 \times R \times B$, where B is the symbol rate. The maximum net bitrate was 557.4 Gb/s at the symbol rate of 216 GBd.

Finally, the transmission performance of the 216-GBd PAM8 signal over 2-km four-core fibre was measured in the O-band experiments. Figure 5(a) shows the NGMI, code rate, and BER for each core. No significant variation in the transmission performances was observed between cores in the experiments. Figure 5(b) shows the achievable and net bitrate for each core after transmission at 526.0, 526.0, 529.2, and 529.2 Gb/s. The total bitrate after 2-km four-core fibre transmission in the O-band was 2.1 Tb/s.

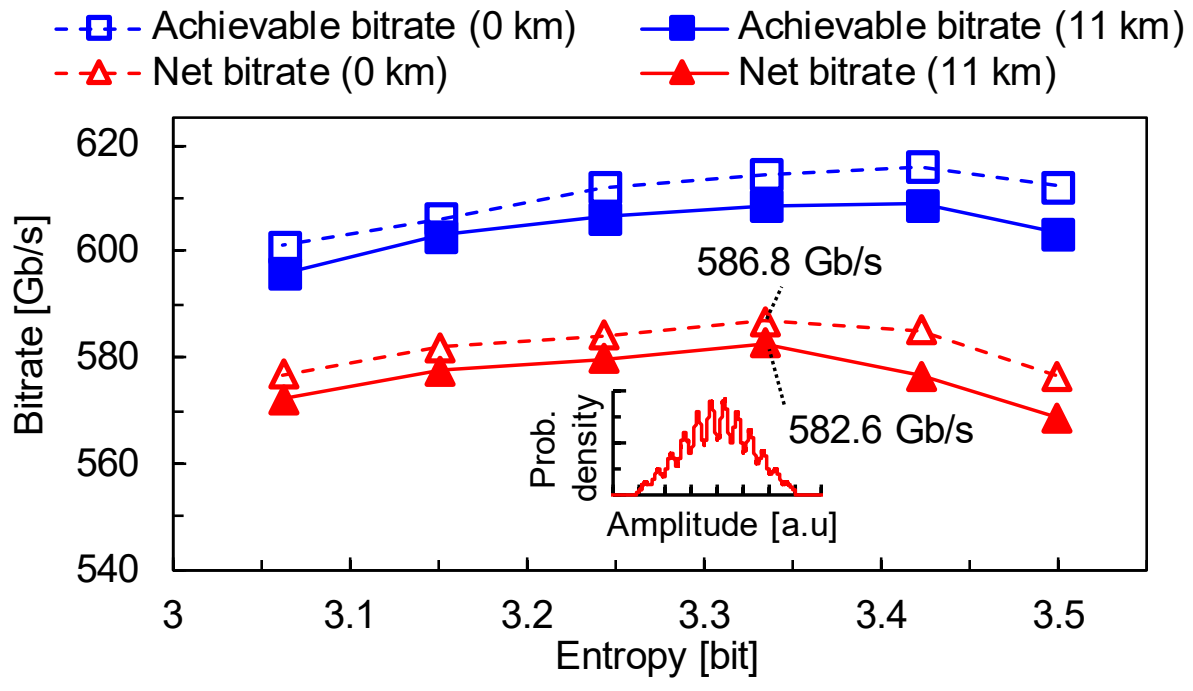

**Fig. 3:** Achievable and net bitrate of 216-GBd PS-PAM12 signal as a function of entropy at back-to-back and after 11-km DSF transmission in C-band experiments.

## Conclusion

In this work, we successfully transmitted a record net bitrate of a net 582-Gb/s PS-PAM12 signal over 11-km DSF in the C-band using a single-carrier 216-GBd IMDD transmission system based on a 150-GHz-BW InP-DHBT-based electrical mixer and a TFLN modulator. We also demonstrated a net 557-Gb/s 216-GBd uniform PAM8 signal generation and detection at back-to-back configuration and 2.1 Tb/s transmission with net 4×526-Gb/s 216-GBd uniform PAM8 signals over 2-km four-core fibre in the O-band. We achieved the highest net bitrate per lane in both the C- and O-bands, thus demonstrating the feasibility of a high-speed IMDD system with a >500 Gb/s/lane solution with >200-GBd signal.

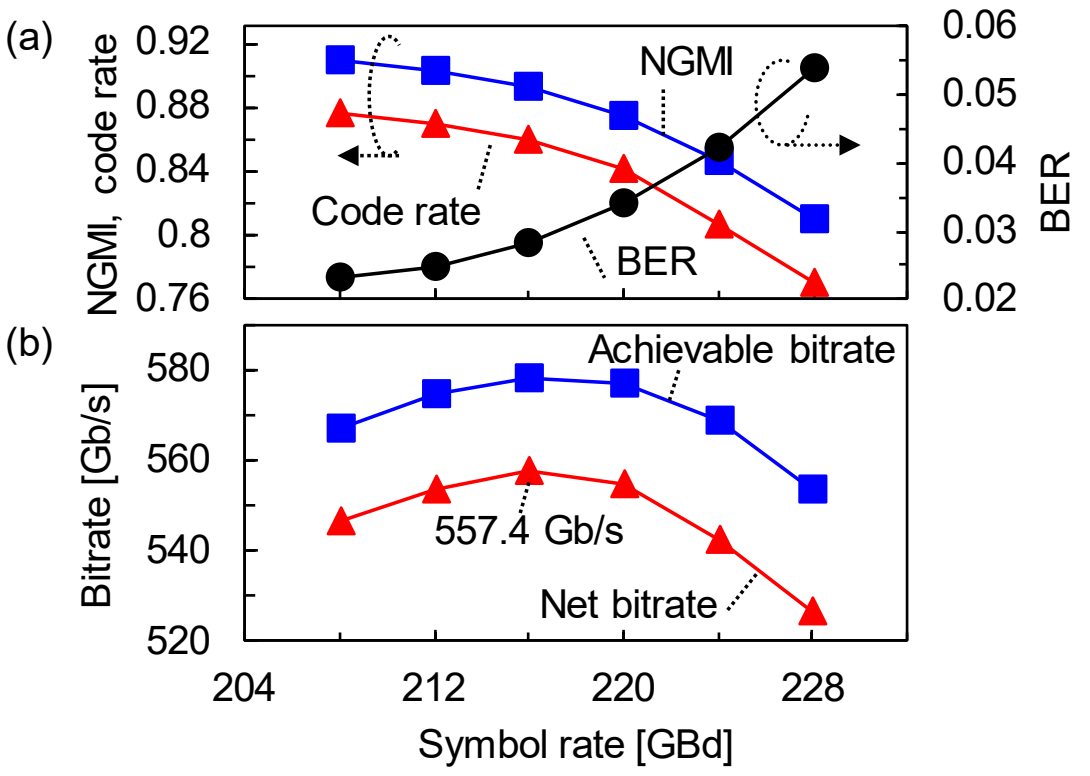

**Fig. 4:** O-band results of uniform PAM8 signal at back-to-back configuration. (a) NGMI, code rate and BER vs. symbol rate. (b) Achievable and net bitrate vs. symbol rate.

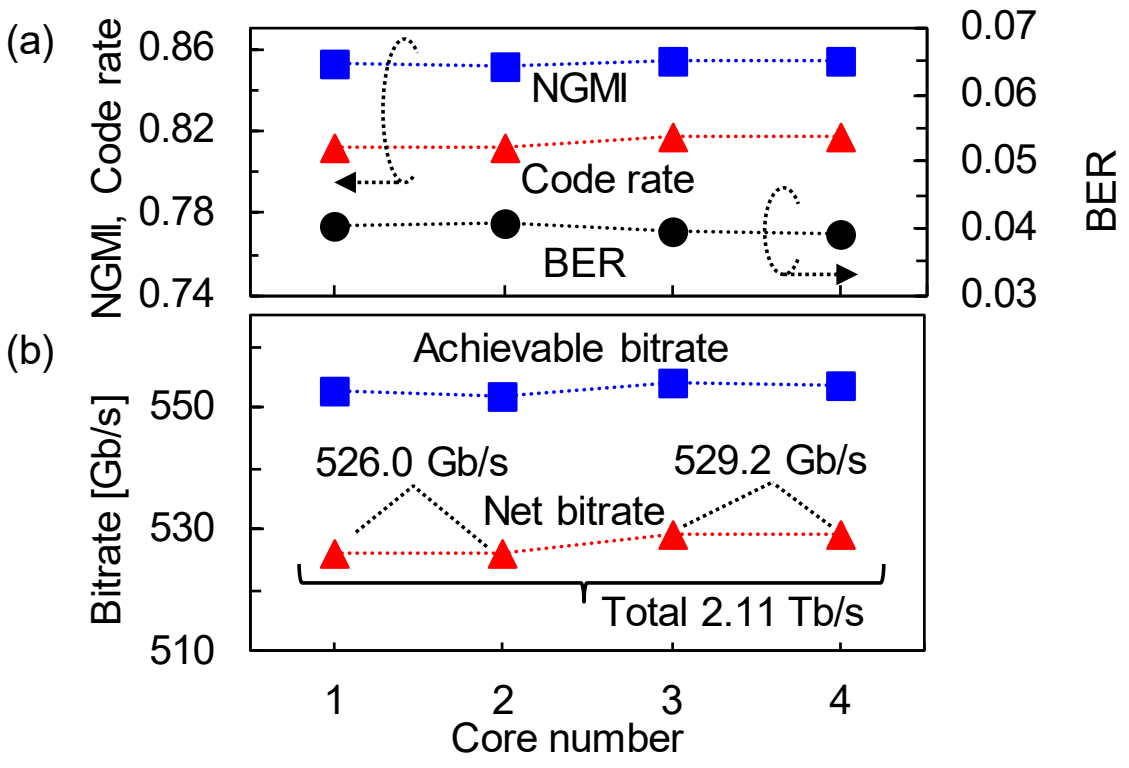

**Fig. 5:** O-band results of 216-GBd uniform PAM8 signal after 2 km transmission. (a) NGMI, code rate and BER for each core. (b) Achievable and net bitrate for each core.